\documentclass[conference,letterpaper]{IEEEtran}

\usepackage[utf8]{inputenc} 
\usepackage[T1]{fontenc}
\usepackage{url}
\usepackage{ifthen}
\usepackage{cite}
\usepackage{svg}
\usepackage[normalem]{ulem}
\usepackage[cmex10]{amsmath} 

\usepackage[hidelinks]{hyperref}
\usepackage{amsmath, epsfig, cite}
\usepackage{amsthm}
\usepackage{amsfonts}
\usepackage{graphicx}
\usepackage{latexsym}
\usepackage{amssymb}
\usepackage{color}
\usepackage{url}
\usepackage{colortbl}
\usepackage{comment}
\usepackage{algorithm}
\usepackage[noend]{algpseudocode}
\usepackage{subcaption}
\usepackage[percent]{overpic}

\theoremstyle{definition}
\newtheorem{theorem}{Theorem}
\newtheorem{lemma}{Lemma}

\newtheorem{definition}{Definition}
\newtheorem{corollary}{Corollary}
\newtheorem{problem}{Problem}
\newtheorem{construction}{Construction}
\newtheorem{example}{Example}

\newcommand{\bfs}{{\boldsymbol s}}

\newcommand{\bfx}{{\boldsymbol x}}
\newcommand{\bfy}{{\boldsymbol y}}

\newcommand{\bfS}{{\mathbf S}}

\newcommand{\cC}{\mathcal{C}}

\newcommand{\cS}{\mathcal{S}}

\newcommand{\cX}{\mathcal{X}}
\newcommand{\cY}{\mathcal{Y}}
\newcommand{\cZ}{\mathcal{Z}}

\renewcommand{\Bbb}{\mathbb}

\newcommand{\N}{{\Bbb N}}

\newcommand{\F}{{\Bbb F}}

\newcommand{\matrixCo}{\Sigma_w^{m\times n}}

\normalsize

\title{\textbf{Error-Correcting Codes for Combinatorial \\ Composite DNA \vspace{-1.25ex}}}
\author{%
    \IEEEauthorblockN{
    \textbf{Omer~Sabary}\IEEEauthorrefmark{1}\IEEEauthorrefmark{2}, 
    \textbf{Inbal~Preuss}\IEEEauthorrefmark{1}\IEEEauthorrefmark{3},  \textbf{Ryan~Gabrys}\IEEEauthorrefmark{4}, \textbf{Zohar~Yakhini}\IEEEauthorrefmark{2}\IEEEauthorrefmark{3}, 
    \textbf{Leon~Anavy}\IEEEauthorrefmark{3}, 
    and \textbf{Eitan~Yaakobi}\IEEEauthorrefmark{2}}
   \IEEEauthorblockA{\IEEEauthorrefmark{2}%
                     Faculty of Computer Science, 
                     Technion---Israel Institute of Technology, 
                     Haifa, Israel}
   \IEEEauthorblockA{\IEEEauthorrefmark{3}%
                     Faculty of Computer Science, 
                     Reichman University, 
                     Hezliya, Israel}
    \IEEEauthorblockA{\IEEEauthorrefmark{4}%
                     Department of Electrical and Computer Engineering, 
                     University of California San Diego, La Jolla, CA, USA}
                      
                     \begin{small}{Emails: \{omersabary, yaakobi\}@cs.technion.ac.il, \{inbalpreuss, leon.anavy\}@gmail.com, rgabrys@eng.ucsd.edu, zohar.yakhini@runi.ac.il}\end{small} 
                     \vspace{-3.5ex}

   \thanks{\IEEEauthorrefmark{1} These authors contributed equally to this work. This project has received funding by the European Union (DiDAX, 101115134). Views and opinions expressed are, however, those of the author(s) only and do not necessarily reflect those of the European Union or the European Research Council Executive Agency. Neither the European Union nor the granting authority can be held responsible for them. This work was also supported in part by NSF Grant CCF2212437.}%
}

\IEEEoverridecommandlockouts

\begin{document} 

\maketitle
\begin{abstract}
Data storage in DNA is developing as a possible solution for archival digital data. Recently, to further increase the potential capacity of DNA-based data storage systems, the combinatorial composite DNA synthesis method was suggested. This approach extends the DNA alphabet by harnessing short DNA fragment reagents, known as \emph{shortmers}. The shortmers are building blocks of the alphabet symbols, each consisting of a fixed number of shortmers. Thus, when information is read, it is possible that one of the shortmers that forms part of the composition of a symbol is missing and therefore the symbol cannot be determined. In this paper, we model this type of error as a type of asymmetric error and propose code constructions that can correct {such errors} in this setup. We also provide a lower bound on the redundancy of such error-correcting codes and give an explicit encoder and decoder for our construction. Our suggested error model is also supported by an analysis of data from actual experiments that produced DNA according to the combinatorial scheme. Lastly, we also provide a statistical evaluation of the probability of observing such error events, as a function of read depth. 
\end{abstract}
\vspace{-.85ex}
\section{Introduction}
\vspace{-.75ex}
In the last decade, there has been notable progress in DNA storage systems, where the stability and density of DNA molecules are utilized to create robust and high-capacity data storage platforms~\cite{YGM16, GBC13, CGK12,BO21}. In standard DNA storage systems, binary data is encoded into sequences over the DNA alphabet $\{A, C, G, T\}$, in which each symbol represents a DNA base (also known as nucleotide). Then, based on these sequences,  DNA molecules called \emph{strands} are generated by a biological process termed \emph{DNA synthesis}, that can only generate multiple copies per strand. The synthesized strands are stored in a storage container.  To read back the binary information the strands are read back into their digital representation using \emph{DNA sequencing}. The sequencing data is called \emph{reads}, and the reads are used as an input to the decoder that retrieves the stored information.  The synthesis, storage, and sequencing are error-prone processes, and thus, to retrieve the information error-correcting codes should be considered.

Recently, Anavy et al, and Choi et al.~\cite{AVA19, CR19} suggested an innovative way to extend the DNA alphabet by harnessing the inherent redundancy (multiple copies) of the synthesis and sequencing process with the use of \emph{composite DNA symbols}. A composite DNA symbol is a representation of a position in a sequence in which there is not just a single base, but a mixture of the four DNA bases. By using a mixture of bases, the alphabet is extended with more symbols that are defined by the bases in the mixture and their ratios. 

Later, Preuss et al.~\cite{PYA21} suggested an extension to the composite DNA synthesis, which is referred to as the \emph{combinatorial composite synthesis method} and {the design of codes suitable for this emergent method} is the main focus of this paper. In the combinatorial composite synthesis method, the building blocks of the composite symbols are the so-called \emph{shortmers}. A shortmer (also known as a \emph{motif}) is a fixed-length sequence that consists of DNA bases. The shortmers are synthesized using a standard DNA synthesis technology, and then they are connected to each other using biochemical process called \emph{ligation}~\cite{PYA21}. In this case, each valid composite symbol is in fact a set of $w \in \N^+$ distinct shortmers. Thus, the alphabet consists of sets of shortmers. To improve the data reliability, and to allow easier detection of the shortmers, they are selected as a subset of all shortmers of a specific length. Other extensions of this method can be found in~\cite{PYA23, RB21}.

The alphabet symbols of the combinatorial composite method are sets of shortmers. Therefore, it is possible that one or more shortmers are not represented in the sequencing reads. In such cases, the observed set of shortmers is missing a subset of them causing an error in reading the data.  In this paper, we model these cases as \emph{asymmetric errors} and study error-correcting codes (ECCs) for the combinatorial composite synthesis. We provide constructions for such codes, including an explicit encoder, and present bounds on their redundancy.  

The rest of this paper is organized as follows. Section~\ref{sec:defs} gives the preliminaries and defines the main problem discussed in the paper. In Section~\ref{sec:cons} we present a construction for a composite asymmetric error-correcting code. In Section~\ref{sec:bounds} we give a sphere-packing bound on such codes. In Section~\ref{sec:enc_dec} explicit encoder and decoder for our construction are provided, while Section~\ref{sec:simulation and stats} presents an analysis of data from previous experiments as well as an evaluation of the probabilities of our discussed error models. Lastly, in Section~\ref{sec: gen e} we give a lower bound on the redundancy of error-correcting codes for a more general error model.



\vspace{-.85ex}
\section{Definitions and Problem Statement} \label{sec:defs}
\vspace{-.5ex}
In the following, we represent our data as a sequence of length $m$ where each element in the sequence is a set of shortmers. Under this representation, each set of shortmers will be represented as a binary vector of length $n$ with exactly $w$ ones in it where the location of the ones in the vector indicate which shortmers are contained in every set. Since every element in our sequence is itself a set, we will find it useful to represent our data as a set of $m \times n$ binary arrays with each row in the array specifying a set of shortmers.

\subsection{Notations}
For a positive integer $n$, let $[n]\triangleq\{0,\ldots,n-1\}$. For a binary vector $\bfx$, $w_H(\bfx)$ denotes the Hamming weight (shortly the weight) of $\bfx$, which is the number of ones in $\bfx$. 

Let $\Sigma = \{A, C, G, T\}$ be the DNA alphabet and let $\ell \in \N$ be the shortmer length. We let $\cS = \{ \bfs_0, \bfs_1, \ldots, \bfs_{n-1} \}$ be a set of $n>1$ \emph{shortmers}, $\bfs_i \in \Sigma^\ell$ for $i\in [n]$, which are indexed lexicographically. For $w<n$, we define the \emph{$w$-combinatorial composite alphabet of $\cS$} as $\Sigma_w^{\cS} \triangleq \{ \bfx_0^{\cS}, \bfx_1^{\cS}, \ldots, \bfx_{\binom{n}{w}-1}^{\cS}\}$, where each \emph{combinatorial composite symbol} $\bfx_i^{\cS}$, for $i\in [\binom{n}{w}]$ is a \emph{set} of $w$ different shortmers chosen from the shortmers set $\cS$. For simplicity, the set of symbols in $\Sigma_w^\cS$ can be abstracted as length-$n$ binary vectors of weight $w$ in which every bit indicates whether a shortmer in $\cS$ belongs to the set. Thus, every $\bfx_i^{\cS} \in \Sigma_w^n$ is mapped to its indicator binary vector, denoted by $\bfx_i\in\{0,1\}^n$ and note that $\sum_{j=0}^{n-1} x_{i, j} = w$. 
From this point, we refer to the composite symbols in our alphabet by their binary vector representation and denote the set of length-$n$ binary vectors of weight $w$ by $\Sigma_w^n$. 

\begin{small}
\begin{example} In ~\cite{PYA21}, the authors used the following parameters {$\ell=3$}, $n=16$ and $w=5$,  

\vspace{-2.5ex}\begin{small}
\begin{align*}\cS =
\begin{Bmatrix}
\bfs_0  = AAT, \bfs_1 = ACA, \bfs_2 = ATG, \bfs_3 = AGC, \\\bfs_4 = TAA, \bfs_5 = TCT, \bfs_6 = TTC, \bfs_7 = TGG, \\ \bfs_8 = GAG, \bfs_{9} = GCC, \bfs_{10} = GTT, \bfs_{11} = GGA, \\ \bfs_{12} = CAC, \bfs_{13} = CCG, \bfs_{14} = CTA, \bfs_{15} = CGT
\end{Bmatrix} .
\end{align*}
\end{small}


The set $\cS$ was selected as a code with Hamming distance of  $d=2$. In this setup, an example of the $99$-th composite symbol of the alphabet is $\bfx_{99}^\cS = (1, 1, 0, 1, 0, 0, 1, 1, 0, 0, 0, 0, 0, 0, 0, 0)$, which represents the set consisting of the shortmers $\{ \bfs_0, \bfs_1, \bfs_3, \bfs_6, \bfs_7\}$. 
\end{example}
\end{small}



A sequence of length $m$ over a composite alphabet $\Sigma_w^\cS$ is denoted by $\cX^\cS = (\bfx_{i_0}^\cS,\ldots,\bfx_{i_{m-1}}^\cS) \in (\Sigma_w^\cS)^m$. This sequence can be abstracted as an $m \times n$ binary matrix $\cX$, in which each row is matched with its corresponding composite symbol from $\Sigma_w^\cS$. That is, 

\vspace{-1.5ex}\begin{small}
$$ \cX =      \begin{pmatrix} \bfx_{i_0} \\ \vdots \\ \bfx_{i_{m-1}} \end{pmatrix} \; = \begin{pmatrix} x_{i_0,0}, x_{i_0,1}, x_{i_0,2} \ldots, x_{i_0,n-1} \\ \vdots \\ x_{i_{m-1},0}, x_{i_{m-1},1}, \ldots, x_{i_{m-1},n-1} \end{pmatrix}, 
$$
\end{small}

\vspace{-.5ex}
\hspace{-2.0ex}and note that for any $h \in [m]$, $\sum_{j=0}^{n-1} x_{i_h,j} = w$. 


In this paper, we consider \emph{the combinatorial composite-DNA channel},
which receives an $m \times n$ matrix $\cX$, and outputs a noisy version of $\cX$, denoted by $\cY$. Similarly, we denote the rows of the matrix $\cY$ by $\bfy_h$, $h \in [m]$, such that $\bfy_h$ is a noisy version of $\bfx_{i_h}$

\vspace{-1.5ex}\begin{small}
$$ \cY =      \begin{pmatrix} \bfy_0 \\ \vdots \\ \bfy_{m-1} \end{pmatrix} \; = \begin{pmatrix} y_{0,0}, y_{0,1}, \ldots, y_{0,n-1} \\ 
\vdots \\ 
y_{m-1,0}, y_{m-1,1}, \ldots, y_{m-1,n-1}  
\end{pmatrix}. 
$$
\end{small}
Lastly, since the exact shortmers in the set $\cS$ do not matter but only the number of shortmers, we refer to the combinatorial composite alphabet from now on by the set $\Sigma_w^n$ and a length-$m$ sequence is simply an $m\times n$ matrix in $\matrixCo$, where  $\matrixCo$ refers to the set of all $m\times n$ matrices in which the weight of every row is $w$.

\subsection{Problem Statement}
Next, we define \emph{composite-asymmetric errors}, which is our main interest in this paper. 

\begin{definition} \label{def:1}
\textbf{Composite asymmetric errors.} For a positive integer $e$ and a row vector $\bfx_i = (x_0, \ldots, x_{n-1}) \in \Sigma_w^n$, we say that the corresponding channel output $\bfy_i = (y_0, \ldots, y_{n-1}) \in \Sigma_{w-e}^n $, suffers from \emph{$e$ composite-asymmetric errors} if $y_i \le x_i$, and $\sum_{i=0}^{n-1} y_i = w-e.$
\end{definition}

Definition~\ref{def:1} can be extended to  matrices as described below. 

\begin{definition}
\textbf{$(t,e)$-composite asymmetric errors.} For positive integers $e$ and $t$ and a matrix $\cX =      \left( \bfx_0,  \ldots  ,\bfx_{m-1} \right)^T \in \Sigma_w^{m\times n}$, we say that the channel output matrix $\cY =       \left( \bfy_0, \ldots, \bfy_{m-1} \right)^T, $ suffers from \emph{$(t,e)$-composite asymmetric errors} if at most $t$ rows of $\cX$ are noisy, each of them suffers from at most $e$ composite-asymmetric errors.
\end{definition}

A \emph{length-$m$ $(n,w)$-composite code $\cC$} is a set of matrices over $\Sigma_w^{m\times n}$ and every codeword in $\cC$ is referred {to} as a \emph{composite codeword}. We say that a length-$m$ $(n,w)$-composite code is a \emph{$(t,e)$-composite-asymmetric ECC (in short $(t,e)$-CAECC)}, if it can correct any $(t,e)$-composite asymmetric error. Such a code will be referred as an \emph{$[m,(n,w);t,e]$-composite code}.


We denote by $A(m,n,w)$ the size of the set of all binary matrices of dimension $m \times n$, in which each row is of weight exactly $w$, that is  $A(m, n, w) =  |\Sigma_w^{m \times n} |= \binom{n}{w}^m$. We denote by $A(m, n, w; t, e)$ the size of the largest $[m,(n,w);t,e]$-composite code. For a composite code $\cC \subseteq \Sigma_w^{m\times n}$, we define its redundancy to be $r(\cC) \triangleq \log(|A(m,n,w)|) - \log(|\cC|)$. Furthermore, we denote by $r(m,n,w; t, e)$ the minimum redundancy of such a composite code. 

The main goal of this paper is to study $(t,e)$-CAECCs and more specifically to solve the following problem. 
\begin{problem}
    Find the value of $A(m,n,w; t, e)$, the size of the largest $[m,(n,w);t,e]$-composite code and correspondingly find the minimum redundancy $r(n, w, m; t, e)$.
\end{problem}



{Although the problem of coding for the asymmetric channel has been studied extensively in the past, our setup departs from previous works such as \cite{BB00,GSSSZ15,Klove95,VT_code} in that the sequences over which our codes are being developed over satisfy a local weight constraint. In particular, recall that each row of our codeword matrices has exactly $w$ ones in it, and this extra information can be leveraged to dramatically reduce the redundancy of our resulting coding schemes. To the best of our knowledge this setup has not been studied before, and one of the goals in this work will be to identify parameter regimes where we can design efficient codes capable of correcting such errors that are larger than traditional asymmetric error-correcting codes. }

\vspace{-.75ex} \section{Code Constructions} \label{sec:cons} \vspace{-.75ex}

In this section we give a code construction for $(t,e)$-CAECCs. 
For any length-$n$ binary vector $\bfx = (x_0, x_1, \ldots, x_{n-1}) \in \{0,1\}^n$, positive integers $\ell$ and $p$, we define the \emph{$\ell$-VT-syndrome over $p$}~\cite{VT_code} of $\bfx$, denoted by $\bfs_{\ell}^p(\bfx)$, as follows 
$ \bfs_{\ell}^p(\bfx) \triangleq (\sum_{i=0}^{n-1} i^\ell x_i) \bmod p.$ 
Note that the VT-syndrome is usually defined such that $\bfs_{\ell}^{n+1}(\bfx) = (\sum_{i=0}^{n-1} i^\ell x_i) \bmod (n+1)$ and therefore $\bfs_{\ell}^{n+1}(\bfx) \in [n+1]$, however, we use the above definition with a prime number to correct multiple errors and to be able to construct outer codes using tensor product codes~\cite{TP_codes}. Therefore, according to our definition, we have that $\bfs_{\ell}^p(\bfx) \in \F_p$, and for the rest of the paper, it is assumed that $p$ is the smallest prime such that $p\ge n$ and according to Bertrand's Postulate~\cite{B1845}, we assume that $n\leq p\leq 2n$.

We also define the \emph{$e$-complete-VT-syndrome over $p$}, denoted by $\bfS_{e}^p(\bfx)$, to be
$ \bfS_{e}^p(\bfx) \triangleq (\bfs_{1}^p(\bfx),\bfs_{2}^p(\bfx),\ldots,\bfs_{e}^p(\bfx)),$ 
and note that $\bfS_{e}^p(\bfx)$ can be interpreted as an element in $\F_{p^e}$. We extend this definition to matrices $\cX \in \matrixCo$, whose rows are given by $\bfx_0, \bfx_1, \ldots, \bfx_{m-1}$, and define the \emph{$e$-complete-VT-syndrome-vector over $p$} of $\cX$, denoted by $\bfS_e^p(\cX)$, to be the vector in which its $i$-th entry corresponds to the $e$-complete-VT-syndrome of the $i$-th row in $\cX$. That is, 
\begin{align*} 
\bfS_e^p(\cX) & \triangleq (\bfS_e^p(\bfx_0), \bfS_e^p(\bfx_1), \ldots, \bfS_e^p(\bfx_{m-1})) \in (\F_{p^e})^{m}. 
\end{align*}

Next, a construction of an $[m,(n,w);t,e]$-composite code is presented. 
\begin{construction} \label{Construction:(t,e)}
Let $e\ge 1$, $p$ be the smallest prime number such that $p\ge n$, and $\cC_t$ be an $[m, k, t+1]$ code over $\F_{p^e}$ capable of correcting $t$ erasures. If $m\le p^e$, then $\cC_t$ is selected as an MDS code with $k=m-t$. The code $\cC_{m, n, w}^{(t,e)}$ is defined as follows. 
 $\cC_{m, n, w}^{(t,e)} =  \{ \cX \in \matrixCo : \bfS_e^p \left( \cX \right)\in \cC_t \}.$

\end{construction}

\begin{theorem} \label{th:construction1}
The code $\cC_{m, n, w}^{(t,e)}$ is a  $(t,e)$-CAECC.  
\end{theorem}
\begin{proof}
Let $\cY$ be an $m\times n$ matrix, obtained from a composite codeword $\cX$ in the code $\cC_{m, n, w}^{(t,e)}$. We denote by $1 \le i_1, \ldots, i_t \le m$, the $t$ indices of the rows of $\cY$ that suffer from $e$ composite-asymmetric errors (if any of these rows suffer from less errors, the proof can be adapted accordingly). That is, 
$w_H(\bfy_{i_1}) = w_H(\bfy_{i_2})= \cdots  = w_H(\bfy_{i_t}) = w-e.$ We prove that it is possible to decode the codeword by proving that any of the above rows can be uniquely decoded.
Without the loss of generality, we show the latter for $\bfy_{i_1}$, and the same proof works for any of the other erroneous rows. As $\cC_t$ is capable of correcting $t$ erasures, by using the decoder of the code $\cC_t$, it is possible to decode the correct $e$-complete-VT-syndrome of the $i_1$-th row, $\bfS_e^p(\bfx_{i_1})$, and thus also the $\ell$-VT-syndrome, $\bfs_\ell^p(\bfx_{i_1})$, for $1 \le \ell \le e$.
We let $h_1< \cdots<h_e$ be the set of $e$ indices corresponding to the locations in which $\bfy_{i_1}$ had asymmetric errors. 
Then we have that for any $1 \le \ell \le e$, 
\vspace{-.8ex}$$h_1^{\ell}+\cdots+h_e^{\ell}  \equiv (\bfs_{\ell}^p(\bfx_{i_1}) - \bfs_{\ell}^p(\bfy_{i_1})) \bmod p.$$\vspace{-.25ex}
Hence, we get $e$ equations per erroneous row. In Theorem 1 in~\cite{D10}, it was shown that these equations have an equivalent polynomial form in which the roots are the indices $h_\ell$, $1 \le \ell \le e$. This polynomial form can be obtained by considering the Newton-Girard formulas~\cite{SO00}. Thus, as was done in~\cite{D10}, Vieta's formula can be used to get the set of roots of the polynomial, which is known to be unique by Lagrange theorem~\cite{N10}. This proves the theorem. 
\end{proof}


Construction~\ref{Construction:(t,e)} leads to the following corollary. 
\vspace{-1ex}\begin{corollary} \label{cor:red_construction_1}
For $m \le p^e$, it holds that 
$r(n, w, m; t, e) \le e t \log (p) \leq e t \log (2n),$
and for $e=1$, we have that, 
$r(m, n, w; t, 1) \le  t \ \log (n).$
\end{corollary}

\begin{proof}
The proof follows by applying Construction~\ref{Construction:(t,e)} with $\cC_t$ chosen as an $[m,m-t]$ MDS code over $\mathbb{F}_{p^e}$. Furthermore, we use all the $p^{et}$ cosets of the code $\cC_t$, which form a partition of the space $\F_{p^e}^m$. Denote these codes by $\cC_t^i$ for $i\in[p^{et}]$. For every code $\cC_t^i$, we construct the corresponding $[m,(n,w);t,e]$-composite code according to Construction~\ref{Construction:(t,e)}, which we denote by $\cC_{m, n, w}^{(t,e),i}$. Since the codes $\cC_t^i$ form a partition of the space $\F_{p^e}^m$, the codes $\cC_{m, n, w}^{(t,e),i}$ form a partition of the space $\matrixCo$, that is, 
\begin{align*}
\left| \dot{\bigcup}_{i\in[p^{et}]} {\cC_{m, n, w}^{(t,e),i}} \right| 
& = \left| \left\{ \cX\in\matrixCo \right\}  \right| =  \binom{n}{w}^m.
\end{align*}
Therefore, by the pigeonhole principle, there exists a code ${\cC_{m, n, w}^{(t,e),i}}$ for some $i\in[p^t]$, such that, 
$ \left| {\cC_{m, n, w}^{(t,e),i}} \right| \ge \frac{\binom{n}{w}^m}{p^{et}}.$
Finally, we have that, 
\begin{align*}
R(m, n, w; &t, e)  \le R({\cC_{m, n, w}^{(t,e),i}}) 
\\ & \le  \log_2\left(\binom{n}{w}^m\right) - \log_2\left(\binom{n}{w}^m\right) + \log_2(p^{et}) )
\\ & = et \log (p).
\end{align*}

\end{proof}


\section{Bounds on the Size of Composite Asymmetric Error-Correcting Codes} \label{sec:bounds}
This section provides a sphere packing bound on the size of $[m, (n,w); t, e]$ composite code. 
We first note that our code $\cC$ is defined over the space $\matrixCo$. Hence, given a codeword $\cX\in\cC$,  any asymmetric error changes the weight of at least one of $\cX$'s rows. Thus, all resulting matrices do not necessarily have the same structure, and therefore the sphere-packing bound cannot be used directly. However, in this section we show how by defining a specific distance and proving some properties on CAECCs, it is indeed possible to get a sphere packing bound for a $(t,e)$-CAECC for any $t$ and $e$.

First, we define the  Hamming distance between two vectors $\bfx, \bfy$ of the same length, denoted by $d_H(\bfx, \bfy)$ as the number of positions in which their bits are different.  Next, we define the $e$-Hamming distance of two matrices $\cX, \cY \in \matrixCo$.

\begin{definition}
    Let $\cX, \cY \in \matrixCo$, and integer $e\ge0 $. Then, 
    \begin{small}
    $$d_{e\textrm{-}H} (\cX, \cY) \triangleq \begin{cases}
    			\infty,  \text{if $\exists i \in [n] : d_H(\bfx_i, \bfy_i) >e$,}\\
            |\{i: \bfx_i \ne \bfy_i \}|,  \text{otherwise.}
    \end{cases}$$
    \end{small}
\end{definition}

The \emph{$e$-Hamming distance} of a code $\cC$, denoted by $d_{e\textrm{-}H}(\cC)$, is defined as the minimum $e$-Hamming distance between any two different codewords in $\cC$. That is, $d_{e\textrm{-}H}(\cC) \triangleq \min_{\cX, \cY \in \cC,\cX\neq \cY}\{ d_{e\textrm{-}H} (\cX, \cY) \}$.
Finally, we define the \emph{$e$-Hamming error ball of radius $t$} of a word $\cX \in \matrixCo$, denoted by $B_{e\textrm{-}H}(\cX,t)$,  as the set of all words in $\matrixCo$ that have $e$-Hamming distance of $t$ or less from $\cX$
$$B_{e\textrm{-}H}(\cX, t) = \{ \cY \in \matrixCo: d_{e\textrm{-}H} (\cX, \cY) \le t \}.$$

\begin{lemma} \label{lm:metric} 
For any two integers $t\le m, e\le w$, it holds that a code $\cC\subseteq \matrixCo $ is a $(t, e)$-CAECC if and only if $d_{2e\textrm{-}H} (\cC)\ge t+1$. 
\end{lemma}
\begin{proof}
First, we assume to the contrary that there exist $\cX, \cY \in \cC$, such that $d_{2e\textrm{-}H} (\cX, \cY) \le t$ and we show that this implies that $\cC$ is not a $(t,e)$-CAECC. From the assumption, there are $t' \le t$ rows, whose indices are denoted by $i_1, i_2, \ldots, i_{t'}$, such that $\cX$ and $\cY$ differ, while the rest of the rows are exactly the same in $\cX$ and $\cY$. 
For any $1\leq h\leq t'$, the $i_h$-th row satisfies, $$1\le d_H(\bfx_{i_h}, \bfy_{i_h}) \le 2e.$$
This implies that for any $1\leq h\leq t'$, the $i_h$-th row of $\cX$ includes $e' \le e$ bits with value of  $1$, while their value in $\cY$ is $0$ and vice versa. Let us consider the word obtained from $\cX$ by flipping these ones into zeros. Similarly, consider the word obtained by flipping the corresponding ones in $\cY$ into zeros as well. The resulting word is the same word, and can be obtained from $\cX$ and from $\cY$ by introducing $e$ asymmetric errors in $t' \le t$ rows, which contradicts $\cC$ being a $(t,e)$-CAECC. 

Next, we assume $d_{2e\textrm{-}H} (\cC) >t$ and we show that $\cC$ is a $(t,e)$-CAECC. From the definition of the code distance, we have that for any $\cX, \cY \in \cC$,  $d_{2e\textrm{-}H} (\cX, \cY) >t$. If $d_{2e\textrm{-}H} (\cC) = \infty$, from the definition of $d_{2e\textrm{-}H}$, there is a some row $i$ such that $d_H(\bfx_i,\bfy_i)>2e$ so any error matrices of $\cX$ and $\cY$ are necessarily distinct. 
Otherwise, we have that $d_{2e\textrm{-}H} (\cC) < \infty$, and for simplicity we assume $d_{2e\textrm{-}H} (\cC) = t+1$. This implies that there are some $t+1$ rows, whose indices are given by $i_1, i_2, \ldots, i_{t+1}$, such that for any $1\leq h \leq t+1$, $1 \le d_{H} (\bfx_{i_h}, \bfy_{i_h}) \le 2e$, i.e., there are different. Since every row has Hamming weight $w$ and only ones can change to zeros, it is necessary to have errors in each of these $t+1$ rows in order to have the same output for $\cX$ and $\cY$. Hence, $\cC$ is a $(t, e)$-CAECC. 
\end{proof}
 
Lemma~\ref{lm:metric} states that a code $\cC\subseteq \matrixCo$ is a $(t,e)$-CAECC if and only if its $2e$-Hamming distance is at least $t+1$. 
The above implies that the radius-$\left\lfloor\frac{t}{2} \right\rfloor$ $2e$-Hamming error balls of $\cC$ are mutually disjoint. Based on this observation, we can compute an explicit sphere-packing bound on $(t,e)$-CAECCs. 

\begin{theorem}\label{th:bound1}
It holds that, 
$$ A(m,n,w;t,e)  \le  \frac{\binom{n}{w}^m}{\left(\frac{m}{\left\lfloor\frac{t}{2} \right\rfloor}\right)^{\left\lfloor\frac{t}{2} \right\rfloor} \cdot \left(\frac{4\cdot w(n-w)}{e^2}\right)^{\frac{e}{2} \cdot \left\lfloor\frac{t}{2} \right\rfloor}}, \text{ and }$$
 
 \vspace{-.85ex}\begin{small}
 \begin{align*}
 r(m, n, w; &t, e)   \ge  \left\lfloor\frac{t}{2} \right\rfloor \log (m) -\left\lfloor\frac{t}{2} \right\rfloor \log \left(\left\lfloor\frac{t}{2} \right\rfloor\right) \\
& \ \ \ + \frac{e}{2} \left\lfloor\frac{t}{2} \right\rfloor \log(w(n-w)) +2\frac{e}{2} \left\lfloor\frac{t}{2} \right\rfloor  -e \left\lfloor\frac{t}{2} \right\rfloor \log(e).
 \end{align*}
 \end{small}
\end{theorem}
\begin{proof} 
We prove the theorem by evaluating the size of the $e$-Hamming error balls of radius $\lfloor \frac{t}{2} \rfloor$, over the space of $\matrixCo$. From the proof of Lemma~\ref{lm:metric}, we have that a code $\cC \subseteq \matrixCo$ is a $(t,e)$-CAECC, if and only if $d_{2e-H}(\cC) \ge t+1$. It can be verified that this implies that for any $\cX, \cY \in \cC$, $B_{e\textrm{-}H}(\cX, \lfloor\frac{t}{2}\rfloor) \cap B_{e\textrm{-}H}(\cY, \lfloor\frac{t}{2}\rfloor) = \emptyset$. Therefore, the maximal size of a $(t,e)$-CAECC can be obtained by evaluating the size of the balls $B_{e\textrm{-}H}(\cX, \lfloor\frac{t}{2}\rfloor)$, which is defined by the selection of 
at most $\lfloor \frac{t}{2} \rfloor$ rows (out of the $m$ given rows), and the selection of at most $e$ bit-flips out of the $n$ bits in each of these rows. Note that, to keep the weight of obtained words to be $w$, there are exactly $\frac{e}{2}$ bit-flips from $0$ to $1$, and exactly $\frac{e}{2}$ bit-flips from $1$ to $0$. 
Thus, we have that, 
\begin{align*}
    A(m,n,w;t,e)  & \le 
    \frac{|\matrixCo|}{\binom{m}{\lfloor \frac{t}{2} \rfloor} \cdot \binom{w}{\frac{e}{2}}^{\lfloor \frac{t}{2} \rfloor } \binom{n-w}{\frac{e}{2}}^{\lfloor \frac{t}{2}  \rfloor} }
    \\ & \le  \frac{\binom{n}{w}^m}{\left(\frac{m}{\left\lfloor\frac{t}{2} \right\rfloor}\right)^{\left\lfloor\frac{t}{2} \right\rfloor} \cdot \left(\frac{4\cdot w(n-w)}{e^2}\right)^{\frac{e}{2} \cdot \left\lfloor\frac{t}{2} \right\rfloor}}.
\end{align*}
Thus, we have that, 
\begin{small}
\begin{align*}
 r(m, n, w;& t, e) \ge \log\left(\left(\frac{m}{\left\lfloor\frac{t}{2} \right\rfloor}\right)^{\left\lfloor\frac{t}{2} \right\rfloor}\cdot \left(\frac{4\cdot w(n-w)}{e^2}\right)^{\frac{e}{2} \cdot \left\lfloor\frac{t}{2} \right\rfloor}\right)
\\ & = \left\lfloor\frac{t}{2} \right\rfloor \log\left(\frac{m}{\left\lfloor\frac{t}{2} \right\rfloor}\right) + \frac{e}{2} \left\lfloor\frac{t}{2} \right\rfloor \log\left(\frac{4w(n-w)}{e^2}\right) 
\\ & = \left\lfloor\frac{t}{2} \right\rfloor \log (m) -\left\lfloor\frac{t}{2} \right\rfloor \log \left(\left\lfloor\frac{t}{2} \right\rfloor\right) \\
& \ \ \ + \frac{e}{2} \left\lfloor\frac{t}{2} \right\rfloor \log(w(n-w)) +2\frac{e}{2} \left\lfloor\frac{t}{2} \right\rfloor  -e \left\lfloor\frac{t}{2} \right\rfloor \log(e).
\end{align*}
\end{small}
\end{proof}

Lastly, we have another bound on $A(m,n,w;t,e)$.
\begin{theorem} \label{th:bound_eitan}
It holds that, 
$ A(m,n,w;t,e) \le \frac{\binom{n}{w}^m}{\binom{n-w+e}{e}^t},$
and $r(m, n, w; t, e) \ge t\log\binom{n-w+e}{e}$.
\end{theorem}
\begin{proof}
    Assume all the errors occurred in the first $t$ rows and exactly $e$ errors in each row. Then, we receive an array in which the weight of each of the first $t$ rows is exactly $w-e$. The number of such arrays is $ \binom{n}{w-e}^t \binom{n}{w}^{m-t}$. Thus, we can apply a sphere packing bound argument considering only these $t$ rows.
    $$ A(m,n,w;t,e) \le \frac{\binom{n}{w-e}^t}{\binom{w}{e}^t} \binom{n}{w}^{m-t}$$
    which implies the bound in the theorem statement.
\end{proof}

Note that the last bound does not depend on $m$, and thus, it will be effective only for smaller values of $m$, while for larger values, it is preferable to apply the bound from Theorem~\ref{th:bound1}.

To give some intuition to our results, let us assume the simplified case in which $m=n=p$ is a prime number, and $w=\frac{n+1}{2}$. In this case, Construction~\ref{Construction:(t,e)} and Corollary~\ref{cor:red_construction_1} show that there exists, for example a $(2,2)$-CAECC with redundancy which is at most $4\log(n)$, while Theorem~\ref{th:bound1} and Theorem~\ref{th:bound_eitan} show that the redundancy is at least $\log(n) + \log(n^2-1)-2$ and $2\log(n^2+n+3)-6$, respectively. Thus, our construction is only constant away from optimality.

\section{Explicit Encoder and Decoder} \label{sec:enc_dec}

\vspace{-.5ex}Next, we describe explicit encoder and decoder for the code $\cC_{m, n, w}^{(t,1)}$ when $n=p$ is a prime number, and $m\le n$.   The overall redundancy of the code is $t\log(n)$. In the next algorithm, we show how to encode $m\cdot \lfloor{\log\binom{n}{w}} \rfloor - t\log(n)$ information bits into a codeword in $\cC_{m, n, w}^{(t,1)}$. Recall that under these given parameters, the code $\cC_{m, n, w}^{(t,1)}$ is defined 
as all the matrices $\cX \in \matrixCo$, such that $\bfS_1^n(\cX) \in \cC_t$, where $\cC_t$ is an $[m, m-t]$ MDS code over $\F_n$.   

Before moving forward into the description of the encoding algorithm, we define the \emph{$(i,n,w)$ coset of the VT-syndrome over n}, denoted by $\bfs_1^n (i, n, w)$, as the set of all vectors $\bfx \in \Sigma_w^n$, such that $\bfs_1^n(\bfx) \equiv i \mod n$. 
Theorem~\ref{th:cosets} states that for coprime $w$ and $n$, the cosets $\bfs_1^n (i, n, w)$ for all $i\in \F_n$ have the same size. 

\vspace{-.5ex}\begin{theorem} \label{th:cosets}
Let $n \in \N^+$ and let $w<n$ such that $w$ and $n$ are coprime, we have that for all $i \in [n]$, 
$|\bfs_1^n (i, n, w)| = \frac{\binom{n}{w}}{n}.$  
\end{theorem}
\vspace{-.7ex}
\begin{proof}
First we show that for all $i\in[n]$ it holds that $|\bfs_1^n (i, n, w)| \leq |\bfs_1^n ((i+w)_n, n, w)|$, where $(i+w)_n$ denotes $i+w \mod n$. 
Let $\bfx \in \bfs_1^n (i, n, w)$, from the definition of $\bfs_1^n (i, n, w)$, we have that there are exactly $w$ ones in $\bfx$ and that $\sum_{j=0}^{n-1} j x_j  \equiv i \mod n$. Let us consider the vector $\bfx'$ that can be obtained by shifting cyclically each of the $w$ ones in $\bfx$ one position to the right. Then, we have that $\bfs_1^n(\bfx') = \sum_{j=0}^{n-1} j x'_j \equiv (i+w) \mod n.$ Since $w<n$, we have that $(i+w)\mod n \neq i \mod n$. Using the same procedure, it is possible to transform any vector $\bfx \in \bfs_1^n (i, n, w)$ into a vector $\bfx' \in \bfs_1^n (i+w, n, w)$ and thus $|\bfs_1^n (i+w, n, w)| \ge |\bfs_1^n (i, n, w)|$.

Next, we recall that $w$ and $n$ are coprime, and thus $w$ is not a divider of $n$. This implies that the above procedure can be applied up to $n$ times on each vector $\bfx \in \bfs_1^n (i, n, w)$, while any time results with a vector $\bfx'$ which is from another coset. The latter implies that the size of any coset is greater than the mean, which proves the theorem's statement. 
\end{proof}
Our encoder assumes a mapping function that receives $\lfloor \log\binom{n}{w} \rfloor$ bits and encodes them into vectors over $\Sigma_w^n$. This mapping is denoted by $E_1$. Furthermore, from Theorem~\ref{th:cosets}, we have that since the size of all the cosets $\bfs_1^n (i, n, w)$ is the same, it is possible to create a mapping that receives a pair of a syndrome in $\F_n$ and $\log(\lfloor \frac{\binom{n}{w}}{n} \rfloor)$ bits of information and encodes them into a vector $\bfx \in \Sigma_w^n$. We denote this mapping by $E_2$.
Our suggested encoder is described in Algorithm~\ref{alg:Enc}. 


\begin{small}
\begin{algorithm}
\caption{Encoding Algorithm} \label{alg:Enc}
\begin{algorithmic}[1]

\begin{small}
\Procedure{Encoder}{}  
\State \textbf{Input.} The encoder receives $m\cdot \lfloor \log\binom{n}{w} \rfloor- t\log(n)$ bits and encodes them into a codeword in $\cC_{w, n, m}^{(t,1)}.$ For this purpose the all zero binary matrix $\cX$ of dimension $m\times n$ is initialized.  
\State \textbf{Step 1.} Take the first $(m-t) \cdot \lfloor \log\binom{n}{w} \rfloor $ bits, and encode them using $E_1$ into $(m-t)$ combintorial-composite symbols over $\Sigma_w^n$. Fill the resulted binary vectors in the first $(m-t)$ rows of $\cX$.
\State \textbf{Step II.} Compute the phantom-syndrome vector of the first $m-t$ rows of $\cX$, $\left(\bfs_1^n (\bfx_1),\ldots,\bfs_1^n (\bfx_{m-t}) \right)$.   
\State \textbf{Step III.} Encode the phantom-syndrome vector $\left( \bfs_1^n(\bfx_1), \ldots, \bfs_1^n(\bfx_{m-t})\right)$ using the encoder of the $[m, m-t]$ MDS code $\cC_{t}$.
By the end of this step, we obtained the encoded phantom syndrome vector, $\bfs_1^n(\cX) = \left(\bfs_1^n(\bfx_1), \ldots, \bfs_1^n(\bfx_{m-t}), r_1, \ldots, r_t ) \right).$ The symbols $r_1, \ldots, r_t$ are the redundancy symbols over $[n]$ of the code $\cC_{t}$. The redundancy symbols $r_1, \ldots, r_t$ can be interpreted as syndromes of the last $t$ rows of the matrix, i.e., rows $m-t+1, \ldots, m$. In particular, it holds that $\bfs_1^n(\cX) =\left(\bfs_1^n(\bfx_1), \ldots, \bfs_1^n(\bfx_{m-t}), r_1, \ldots, r_t \right) =  \left(\bfs_1^n(\bfx_1), \ldots, \bfs_1^n(\bfx_{m-t}), \bfs_1^n(\bfx_{m-t+1}), \ldots, \bfs_1^n(\bfx_{m}) \right).$
\State \textbf{Step IV.}. The last $t$ rows of the matrix $\cX$ are encoded as follows. For $1 \le i \le t$, the $m-t+i$-th row of $\cX$ is encoded with $E_2$ by considering the combination of $\bfs_1^n(\bfx_{m-t+i}) = r_{i}$ and $\log(\lfloor \frac{\binom{n}{w}}{n} \rfloor)$  bits of information. 
\State \textbf{Output.} The matrix $\cX$ is returned as output. 
\EndProcedure 
\end{small}
\end{algorithmic} 
\end{algorithm} 
\end{small} 

\vspace{-.75ex}
\section{Simulation and Statistics} \label{sec:simulation and stats}
\vspace{-.8ex} In this section, we analyze data from previous experiments~\cite{PYA23, PYA21} to support our channel model and error characterization. 
Furthermore, we provide an evaluation of the error probabilities of observing asymmetric errors.   

\vspace{-1.25ex}\subsection{Statistics on real data}
\begin{figure*}[htbp]
  \centering

  \begin{subfigure}[b]{0.47\linewidth}
    \centering
    \includegraphics[width=1.15\textwidth]{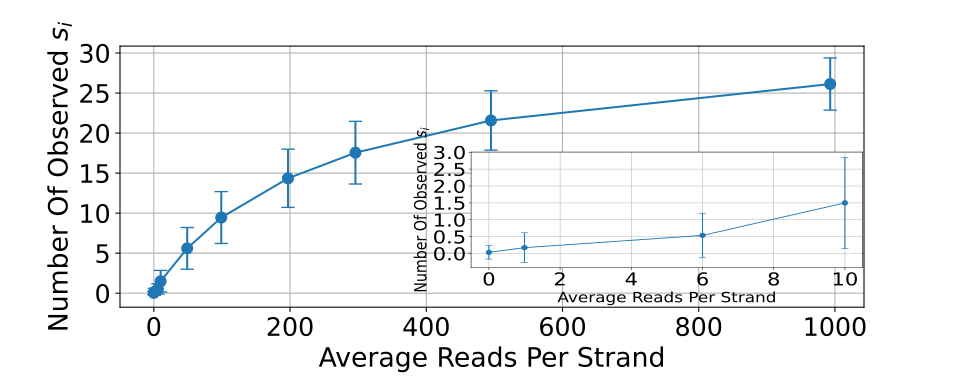}
    \caption{\vspace{-1.3ex}Data of~\cite{PYA23} with $n=96$, $w=32$, $m=1$, 8 different strands.}
    \label{fig:Asymmetric combinatorial errors in experimental results-a}
  \end{subfigure}
  \hfill 
  \begin{subfigure}[b]{0.47\linewidth}
    \centering
    \includegraphics[width=1.15\textwidth]{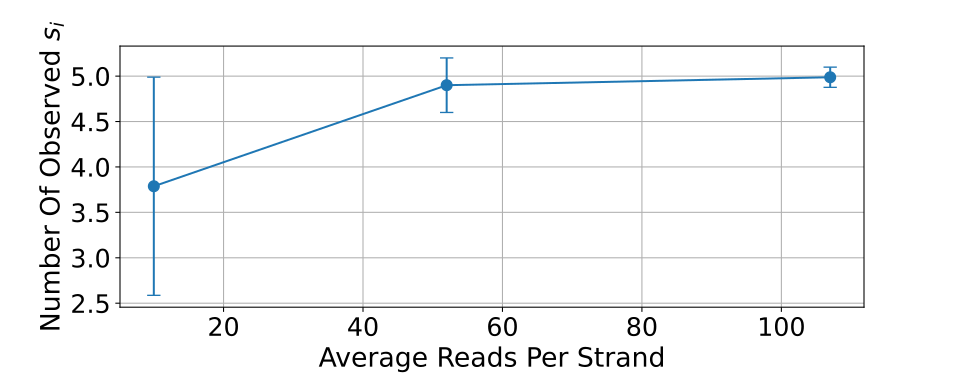}
    \caption{\vspace{-1.3ex}Data of~\cite{PYA21} with $n=16$, $w=5$, $m=4$, 2 different strands.}
    \label{fig:Asymmetric combinatorial errors in experimental results-b}
  \end{subfigure}

  \caption{Asymmetric combinatorial errors in experimental results. The x-axis represents the average reads per strand, in sampling from actual NGS data. The y-axis shows the number of observed $s_{i}$. Midpoints represent the mean count of observed $s_{i}$, and the whiskers represent the std of 10 repeated samplings aggregated over the different strands to each experiment.\vspace{-4ex}}
  \label{fig:Asymmetric combinatorial errors in experimental results}
\end{figure*}

To emphasize the importance of error correction code in recovering data effectively, Fig. \ref{fig:Asymmetric combinatorial errors in experimental results} shows the direct correlation that exists between the quality and quantity of sequencing reads and the number of shortmers observed. An insufficient number of observed shortmers can lead to errors in data recovery, which are defined in this paper as asymmetric errors. Fig. \ref{fig:Asymmetric combinatorial errors in experimental results} depicts data from two combinatorial composite shortmers experiments with different synthesis protocols and different sequencing technologies~\cite{PYA21}~\cite{PYA23}. The plots represent sampling of reads from the overall full set of reads with varying sampling rates. The number of observed unique shortmers is plotted against the average number of reads per strand. Fig. \ref{fig:Asymmetric combinatorial errors in experimental results-a} shows results from ~\cite{PYA23} where a single combinatorial synthesis cycle was demonstrated (meaning $m=1$) with a $n=96$ and $w=32$. The sequencing in this experiment was performed using Oxford Nanopore MinION. Clearly, even with an average coverage of 1,000 reads we could not recover all $32$ shortmers. Fig. \ref{fig:Asymmetric combinatorial errors in experimental results-b} shows the results from~\cite{PYA21} where four combinatorial synthesis cycles ($m=4$) were demonstrated with $n=16$ and $w=5$.  The sequencing was performed using Illumina MiSeq. In this case, a coverage of 100 reads was sufficient for recovering all five shortmers. Note that it is possible that more than five shortmers can be observed due to wrong classification or other  experimental errors.

\vspace{-1ex}\subsection{Evaluation of error probability}

\begin{figure}[htbp]
  \centering
    \centering \vspace{-3.3ex}
    \includegraphics[width=75mm]{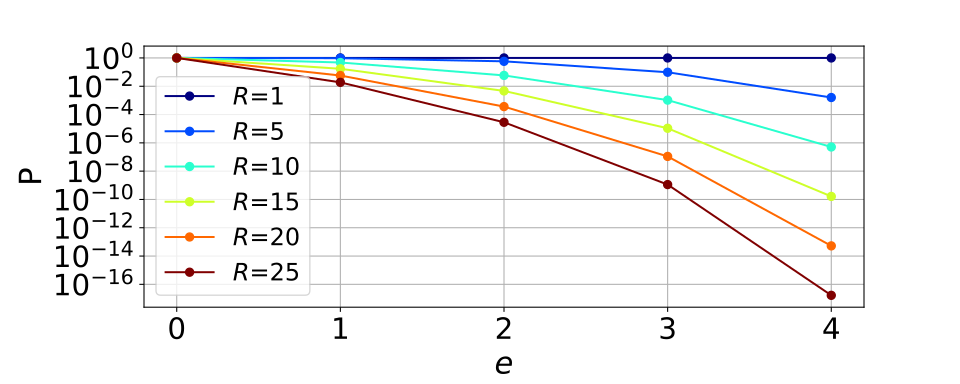} \vspace{-1.5ex}
    \label{fig:Probability to observe e asymmetric combinatorial errors or more in a single combinatorial letter-b}

  \caption{Probability to observe $e$ asymmetric errors or more in a single combinatorial symbols. The x-axis indicates $e$ or more errors, each line represents a different number of analyzed reads ($R$) and the y-axis shows the error probability. Results for $w=5,R=1,5,10,20,25,e=0,1,…,4$.\vspace{-2.0ex}}
  \label{fig:Probability to observe e asymmetric combinatorial errors or more in a single combinatorial letter}
\end{figure}

Fig. \ref{fig:Probability to observe e asymmetric combinatorial errors or more in a single combinatorial letter} and Table \ref{tab:Probability of observing at most asymmetric combinatorial errors} depict the probabilities of observing composite asymmetric errors directly calculated using the coupon collector’s model described in ~\cite{PYA24}. It should be noted that this calculation is based on the combinatorial factor $w$ alone, ignoring $n$, as the model assumes uniform sampling of the $w$ observed $s_{i}$. In Fig. \ref{fig:Probability to observe e asymmetric combinatorial errors or more in a single combinatorial letter} the probability of observing $e$ errors or more is shown as a function of the number of analyzed reads ($R$) for a single combinatorial letter ($m=1$) using combinatorial factor $w=5$. As expected, the error probability decreases as more reads are analyzed. However, it is likely to observe several composite asymmetric errors when analyzing 10-20 reads. This emphasizes the need for efficient ECCs for correcting composite asymmetric errors.

Table \ref{tab:Probability of observing at most asymmetric combinatorial errors} shows direct calculation of the probability of a message encoded using the $(t,e)$ code to be successfully decoded. That is, the table presents the probability of observing at most $t$ letters with at most $e$ errors in each for a combinatorial word of length $m=10$ and combinatorial factor $w=5$. For instance, the probability of observing at most one symbol with at most one error is presented in the first cell ($p(1,1)=0.0137$). Clearly, the probability increases as $t$ and $e$ increase. 

\begin{table}[htbp]
  \centering
  \vspace{-.75ex}\includegraphics[width=70mm]{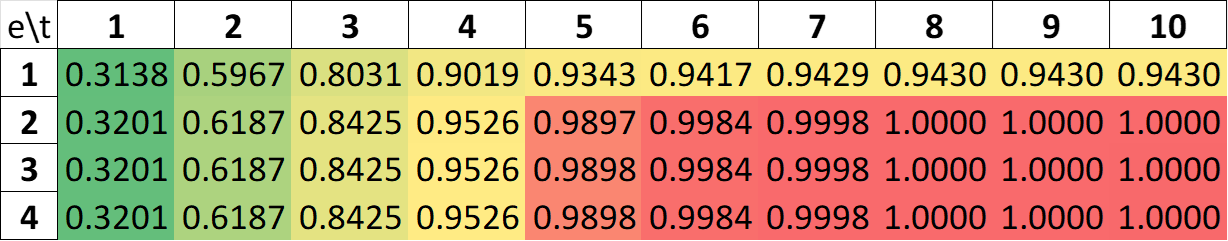}
  \caption{Probability of observing at most $(t,e)$ asymmetric combinatorial errors. Rows correspond to $t$. Columns correspond to $e$. In each calculation $w=4, R=10,m=10,e=1,2,…,5,t=1,2,…,10$ and $n>w$. \vspace{-3ex}}
  \label{tab:Probability of observing at most asymmetric combinatorial errors}
\end{table}

\vspace{-1.05ex} \section{Extensions of the asymmetric error model} \label{sec: gen e} \vspace{-.65ex}
This section discusses more generalized error models that include asymmetric errors. 

\vspace{-1.25ex}\subsection{$(t_1,t_2)$-CAECCs}
In practice, as can be seen in our analysis of data from previous experiments in Section~\ref{sec:simulation and stats},  asymmetric errors of more than $e=2$ shortmers rarely happen. However, it is more likely that $t_1$ rows can suffer from $1$ asymmetric errors each, while $t_2 < t_1$ rows suffer from $e=2$ asymmetric errors. Codes that can correct errors of this pattern are termed $(t_1, t_2)-$CAECC.  
For two integers $t_1 \ge t_2$, we assume $\cC_{t_1+t_2}, \cC_{t_2}$ are codes correcting $t_1+t_2, t_2$ erasures (respectively) over $\F_p$, where $p$ is the smallest prime number, such that $p\ge n$. Construction~\ref{Construction:(t,e)} can be extended to form a $(t_1, t_2)-$CAECC as follows.

\vspace{-.75ex}\begin{construction} \label{Construction:(t1,t2)}
Let $p$ be the smallest prime number such that $n \le p$ and $m\le p$. Then, we have that, 
\vspace{-.5ex}\begin{small}
\begin{align*} 
 &\cC_{(t_1, t_2)} =\{ \cX \in \matrixCo  : \bfS_1^p(\cX) \in \cC_{t_1+t_2}, \bfS_2^p(\cX) \in \cC_{t_2}\}.
\end{align*}
\end{small}

\end{construction}

\vspace{-1ex}Using the same techniques that were used in the proof of Construction~\ref{Construction:(t,e)} it can be shown that the code $\cC_{(t_1, t_2)}$ can correct up to $t_1$ rows with $1$ asymmetric error and $t_2$ rows with $2$ asymmetric errors. 

\vspace{-1.2ex} \begin{corollary} \label{cor:redundancy_t_1_t_2}
    There exists a $(t_1, t_2)$-CAECC with redundancy of 
    $ (t_1+2t_2)\log(p).$
\end{corollary}

\vspace{-.85ex} The latter construction can be further extended to correct  $e_1$ in $t_1$ rows and $e_2 \ge e_1$ errors in $t_2$ rows. 

\vspace{-.65ex}\begin{construction} \label{Construction:(t1,t2)_e}
Let $e_2 \ge e_1 \ge 1$, and let $p$ be the smallest prime number such that $n \le p$. Then, we have that, 
\begin{small}\vspace{-.95ex}
\begin{align*} 
 \cC_{(t_1, e_1, t_2, e_2)} =\{ \cX & \in \matrixCo : \bfS_i^p(\cX) \in \cC_{t_1+t_2}, \bfS_j^p(\cX) \in \cC_{t_2}, \\ & \text{ for any }1 \le i \le e_1 \text{ and } 1 \le j \le e_2\}.
\end{align*}
\end{small} 
\end{construction} \vspace{-.95ex}
Similarly to Theorem~\ref{th:bound_eitan}, we can show that a lower bound on the redundancy of the code from the latter construction is $t_1 \log (\binom{n-w+e_1}{e_1}) +t_2 \log(\binom{n-w+e_2}{e_2}).$
\vspace{-.5ex}\subsection{$2$-CAECC}
\vspace{-.5ex} Lastly, we study the case in which the channel can introduce up to $e=2$ composite asymmetric errors \emph{anywhere} in the codeword $\cX\in \matrixCo$. Such codes are termed \emph{$2$-composite asymmetric error-correcting codes ($2$-CAECC)}. Next, we give a sufficient condition for a code to correct $e=2$ asymmetric errors. For a word $\cX\in\matrixCo$ we define $A_e(\cX)$ as the \emph{$e$-asymmetric error ball} of $\cX$ as all the words that can be obtained from $\cX$ by introducing up to two asymmetric errors.

\vspace{-1.05ex} \begin{lemma} \label{lm:general_e_2}
Let $\cX, \cY \in \matrixCo$ we have that if 
$A_2(\cX) \cap A_2(\cY) \ne \emptyset$
then,
$B_{2\textrm{-}H}(\cX,1) \cap B_{2\textrm{-}H}(\cY,1) \ne \emptyset$.
Hence, given a code $\cC \subseteq \matrixCo$, if $\cC$ satisfies  $B_{2\textrm{-}H}(\cX,1) \cap B_{2\textrm{-}H}(\cY,1) = \emptyset$ for any $\cX, \cY \in \cC$ then $\cC$ is $2$-CAECC.  
\end{lemma}

\begin{proof}
    Let $\cX, \cY \in \matrixCo$ as in the lemma statement such that $A_2(\cX) \cap A_2(\cY) \ne \emptyset$. Let us consider $\cZ \in A_2(\cX) \cap A_2(\cY)$, we now show that there is a word $\cZ' \in B_{2\textrm{-}H}(\cX,1) \cap B_{2\textrm{-}H}(\cY,1)$ which prove the  statement in the lemma. First, since $\cZ$ has one or two rows of weight less than $w$, the rows in which $\cX$ and $\cY$ have asymmetric errors are the same. In particular, either there is a single row with two asymmetric errors in $\cX$ and $\cY$, or there are two different rows with a single asymmetric error in each. Therefore, let us consider the two following cases.
    
    In the first case, the two asymmetric errors are in the same row in $\cX$ and $\cY$, whose index is given by $1\le i \le M$ and let $1\le j_1<j_2 \le n$ be the indices of the columns in $\cX$ that have experienced the asymmetric errors and let $1\le j^{'}_1<j^{'}_2 \le n$ be the corresponding indices of the asymmetric errors in $\cY$. Let us consider the word $\cZ'$ obtained from $\cX$ by flipping the bit in the $(i,j^{'}_2)$-th entry from $0$ to $1$, and the bit in the $(i, j_2)$-th entry from $1$ to $0$. The word $\cZ'$ can be also obtained from $\cY$ by flipping the bit in the $(i,j_1)$-th entry from $0$ to $1$ and the bit in the $(i, j^{'}_1)$-th entry from $1$ to $0$. Thus $\cZ' \in B_{2\textrm{-}H}(\cX,1) \cap B_{2\textrm{-}H}(\cY,1).$

    In the second case, the two asymmetric errors are in two different rows whose indices are given by $i$, and $i'$. 
    In this case, $\cZ$ is obtained from $\cX$ by introducing asymmetric errors in the entries $(i, j_1)$ and  $(i', j_2)$ and from $\cY$ by asymmetric  errors in the entries $(i, j^{'}_1)$ and  $(i', j^{'}_2)$.
    The word $\cZ'$ can be obtained from $\cX$ by flipping the bit in the $(i,j^{'}_1)$-th entry from $0$ to $1$, and the bit in the $(i, j_1)$-th entry from $1$ to $0$. The word $\cZ'$ can be also obtained from $\cY$ by flipping the bit in the $(i',j_2)$-th entry from $0$ to $1$ and the bit in the $(i', j^{'}_2)$-th entry from $1$ to $0$. Thus $\cZ' \in B_{2\textrm{-}H}(\cX,1) \cap B_{2\textrm{-}H}(\cY,1).$
\end{proof}
\vspace{-.95ex}Lemma~\ref{lm:general_e_2} implies that by considering the size of the $e$-Hamming error balls of radius 1, it is possible to apply a sphere packing bound on $2$-CAECCs. For this purpose, we denote by $A(m,n,w,e)$ the maximum size of an $e$-CAECC, where $r(m,n,w,e)$ denotes its minimum redundancy. 

\vspace{-1ex}\begin{theorem} \label{th:sphere_packing_e}
It holds that, 
$A(m,n,w,e) \le \frac{\binom{n}{w}}{mw(n-w)}, $
and 
$r(m,n,w,e) \ge \log(m\cdot (w)\cdot (n-w)).$
\end{theorem}
\begin{proof}
The theorem follows from Lemma~\ref{lm:general_e_2} and by considering the sphere packing bound.  Given a code $\cC$ which is a $2$-CAECC, for any $\cX \in \cC$ the size of $B_{2\textrm{-}H} (\cX, 1)$ is given by selecting the row, selecting the bit that is flipped from $1$ to $0$ and by selecting the bit that is flipped from $0$ to $1$. 
\end{proof}

\vspace{-.75ex}Lastly, it should be noted that using Construction~\ref{Construction:(t1,t2)} it is possible to create a a $2$-CAECC that by  Corollary~\ref{cor:redundancy_t_1_t_2} has a redundancy of $3\log(p)$. Assuming $m=n=p$ prime and $w=\frac{p+1}{2}$, we get by Theorem~\ref{th:sphere_packing_e} that an optimal redundancy of such code is $\log(p\frac{p+1}{2}\frac{p-1}{2})=3\log(p)+\log(1-\frac{1}{p^2})-2 \ge 3\log(p) -2.5 $, which implies that Construction~\ref{Construction:(t1,t2)} is only a constant away from optimality.

\end{document}